\documentclass[proof]{pasj00}
\usepackage[dvips]{color}
\draft

\begin{document}
\SetRunningHead{T. Nishimichi et al.}
{Modeling Nonlinear Evolution of Baryon Acoustic Oscillations}
\Received{????/??/??}
\Accepted{????/??/??}

\title{Modeling Nonlinear Evolution of Baryon Acoustic Oscillations:
Convergence Regime of $N$-body Simulations and Analytic Models}
\author{%
  Takahiro \textsc{Nishimichi}\altaffilmark{1}, Akihito
  \textsc{Shirata}\altaffilmark{1,2}, Atsushi
  \textsc{Taruya}\altaffilmark{1,3,5},
  Kazuhiro \textsc{Yahata}\altaffilmark{1},\\
  Shun \textsc{Saito}\altaffilmark{1}, Yasushi
  \textsc{Suto}\altaffilmark{1,3}, Ryuichi
  \textsc{Takahashi}\altaffilmark{4}, Naoki
  \textsc{Yoshida}\altaffilmark{4,5}, Takahiko
  \textsc{Matsubara}\altaffilmark{4},\\
  Naoshi \textsc{Sugiyama}\altaffilmark{4,5}, Issha
  \textsc{Kayo}\altaffilmark{4,5}, Yipeng \textsc{Jing}\altaffilmark{6},
  and, Kohji \textsc{Yoshikawa}\altaffilmark{7} } 
\altaffiltext{1}
{Department of Physics, School of Science, The University of Tokyo,
  Tokyo 113-0033} 
\altaffiltext{2} {Department of Physics,
  Tokyo Institute of Technology, Tokyo 152-8511}
\altaffiltext{3} {Research Center for the Early Universe, The
  University of Tokyo, Tokyo 113-0033} 
\altaffiltext{4}
{Department of Physics and Astrophysics, Nagoya University, Chikusa,
  Nagoya 464-8602} 
\altaffiltext{5}
{Institute of Physics and Mathematics of the Universe, The University of Tokyo,\\
  5-1-5 Kashiwa-no-ha, Kashiwa City, Chiba 277-8582}
\altaffiltext{6} {
  Shanghai Astronomical Observatory,
  Nandan Road 80, Shanghai 200030, China} 
\altaffiltext{7} {Centre for
  Computational Sciences, University of Tsukuba, Tsukuba 305-8577}
\email{nishimichi@utap.phys.s.u-tokyo.ac.jp}
\maketitle

\begin{abstract}
    We use a series of cosmological $N$-body simulations and various
    analytic models to study the evolution of the matter power
    spectrum in real space in a $\Lambda$ Cold Dark Matter universe.
    We compare the results of $N$-body simulations against three
    analytical model predictions; standard perturbation theory,
    renormalized perturbation theory, and the closure approximation.
    We include the effects from finite simulation box size in the
    comparison. We determine the values of the maximum wavenumbers,
    $k^{\rm lim}_{1\%}$ and $k^{\rm lim}_{3\%}$, below which the
    analytic models and the simulation results agree to within $1$ and
    $3$ percent, respectively. We then provide a simple empirical
    function which describes the convergence regime determined by
    comparison between our simulations and the analytical models. We
    find that if we use the Fourier modes within the convergence
    regime alone, the characteristic scale of baryon acoustic
    oscillations can be determined within $1\%$ accuracy from future
    surveys with a volume of a few $h^{-3}$Gpc$^3$ at $z\sim1$ or
    $z\sim3$ in the absence of any systematic distortion of the power
    spectrum.
\end{abstract}
\section{Introduction}
\label{sec:Intro}
The nature of dark energy remains one of the most fundamental
questions in physics and cosmology.  It is accessible only through
astronomical observations, and a number of large galaxy redshift
surveys and weak-lensing observations are expected to yield tight
constraints on the dark energy equation of state parameter, $w_{\rm
  DE}\equiv p_{\rm DE}/\rho_{\rm DE}$. The use of baryon acoustic
oscillations (hereafter BAOs) is a promising tool to determine $w_{\rm
  DE}$. The characteristic scale of BAOs
can be used as a robust standard ruler, which helps us to reconstruct
the expansion history of the universe.

Substantial improvements in theories as well as observations are
required to constrain $w_{\rm DE}$ within an accuracy of a few
percent, which is the goal of next generation BAO
surveys. \citet{Nishimichi2007} showed that the measurements of the
angular diameter distance and of the Hubble parameter must be very
accurate, with errors less than $1\%$, in order to constrain $w_{\rm
  DE}$ at a $5\%$ level at $z\sim1$ in the special case that the other
cosmological parameters are fixed.  Clearly, theoretical models are
required to predict the matter power spectrum with even better
accuracy.

Recently, significant progress has been made in
developing analytic models of the matter power spectrum
by extending conventional perturbation theory (e.g., Crocce \&
Scoccimarro 2006a, b, 2008; Matarresse \& Pietroni 2007, 2008;
\cite{Valageas2007,Taruya2008,Matsubara2008}).  Cosmological $N$-body
simulations are often used as a reference to calibrate these models.
It is not clear, however, whether or not $N$-body simulations provide
sufficiently accurate results at the required percent
level even at weakly nonlinear scales where perturbation theory is
generally supposed to be reasonably accurate.

Indeed a variety of systematic effects needs to be considered in
interpreting the results of $N$-body simulations. It is well known
that discreteness of a density field in $N$-body simulations causes
spurious behaviour of the power spectrum at length scales comparable
to or smaller than the mean inter-particle separation
(\cite{Melott1997,Splinter1998,Hamana2002,Baertschiger2002,Marcos2006};
Joyce \& Marcos 2007a, b; \cite{Joyce2008,Romeo2008}).  Adopting a
finite-size simulation box and imposing a periodic boundary condition
also systematically biases the growth of density fluctuations at
almost all the length scales \citep{Seto1999}.  Recently,
\citet{Takahashi2008} examined the effect of the finite number of
Fourier modes in a {\it single} $N$-body realization due to a finite
box size. They found that the evolution of the power spectrum measured
from each realization deviates from the prediction of perturbation
theory by more than a few percent even at very large scales (e.g.
$k\lesssim0.1h$Mpc$^{-1}$) and that this deviation results from having
only a finite number of modes, which can be explained by including
second-order perturbation contributions.  This implies that a
sub-percent level accuracy both in theories and in simulations is very
demanding and we need to explore this in detail.

In the present paper, we consider a specialized case of the matter
power spectrum in real space. We critically compare the evolution of
the matter density fluctuations in the weakly nonlinear regime in
cosmological simulations to that of theoretical predictions.  We
measure the matter power spectra using the simulation outputs and
compare them with analytic models. In doing so we correct for the
finite-mode effect pointed out by \citet{Takahashi2008}. Unlike the
conventional approach, we do not regard the results of $N$-body
simulations as perfect calibrators of analytical models. We determine
ranges of wavenumbers where both theories and $N$-body simulations
agree within a given accuracy.

The rest of the present paper is organized as follows. Section
\ref{sec:nonlinear} briefly outlines the analytic models that we
examine for the nonlinear matter power spectrum. Section
\ref{sec:simulations} describes our simulation details. Our methods to
measure the matter power spectrum and to correct for the finite volume
effect are shown in section \ref{sec:power}. We show our results of
$N$-body simulations and determine the convergence regime in
wavenumber in section \ref{sec:results}. We also discuss the phase
information of BAOs and predict the parameter forecast from the modes
in this convergence regime alone in section
\ref{sec:implication}. Finally section \ref{sec:summary} summarizes
the present paper. The convergence tests of our $N$-body simulations
are discussed in the appendix.

\section{Analytic Nonlinear Models}
\label{sec:nonlinear}

Various analytic models have been proposed to account
for nonlinear evolution of the matter power spectra
in real space.
In this paper, we will compare $N$-body simulations with three
different analytic models, specifically focusing on the leading-order
contributions.  The models include standard perturbation theory (SPT;
e.g., \cite{Bernardeau2002}), renormalized perturbation theory (RPT;
\cite{Crocce2006a}a, b, 2008), and the closure
approximation (CLA; \cite{Taruya2008}).

SPT is a straightforward expansion of the fluid
equations around their linear solution, assuming that fluctuation
amplitudes are small. Schematically, the expansion is written
\begin{equation}
P(k,z) = P^{\rm L}(k,z) + P^{\rm 1-loop}(k,z)+\cdots,
\label{eq:spt}
\end{equation}
where $P^{\rm L}(k,z)$ is the linear power spectrum, which grows as
$\propto D_+^2(z)$ with linear growth rate, $D_+(z)$. The second term,
$P^{\rm 1-loop}(k,z)$, is the one-loop correction to the power
spectrum, which represents the contributions coming from the
second-order and third-order solutions. The one-loop correction is
roughly proportional to $P^{\rm L}\Delta_{0}^2$ with
$\Delta_0^2=k^3P^{\rm L}(k,z)/(2\pi^2)$, and this term subsequently
exceeds the linear term at lower redshift and smaller scales. The
explicit expressions for the one-loop correction $P^{\rm 1-loop}(k,z)$
may be found in the literature (e.g.,
\cite{Makino1992,Jain1994,Scoccimarro1996,Jeong2006,Nishimichi2007}).

Both RPT and CLA are constructed from the renormalized expression of a
perturbation series obtained from SPT, by introducing non-perturbative
statistical quantities such as the propagator and the vertex
function. The resultant {\it renormalized} expressions for the power
spectrum are given as an infinite series of irreducible diagrams
composed of the propagator, the vertex function, and the power spectra
themselves in a fully non-perturbative way. Then, employing the Born
approximation, RPT of \citet{Crocce2008} perturbatively evaluates the
renormalized diagrams under the tree-level approximation of vertex
functions. In contrast, CLA proposed by \citet{Taruya2008} first
truncates the renormalized expansion at the one-loop order under the
tree-level approximation of vertex functions, and obtains a closed set
of equations for the power spectrum and the propagator. In practice,
they further apply the Born approximation to the truncated diagram. As
a result, the leading-order calculations of the power spectra in CLA
reduce to the same analytic expression as in the case of RPT, leading
to
\begin{equation}
P(k,z) = G^2(k,z)P^{\rm L}(k,z_{\rm ini}) + P_{\rm MC}^{\rm 1-loop}(k,z)+\cdots.
\label{eq:Pk_RPT_CLA}
\end{equation}
Here, the function $G(k,z)$ is the propagator, and the term $P_{\rm
  MC}^{\rm 1-loop}(k,z)$ represents the corrections generated by
mode-mode coupling at smaller scales, constructed from the one-loop
diagram. In the above expression, the only difference between CLA and
RPT is the asymptotic behavior of the propagator. The explicit
expressions for the propagator together with the
mode-coupling term are summarized in \authorcite{Crocce2006b} (2006b,
2008) and \citet{Taruya2008}.

We note here that the most important difference between SPT and
RPT/CLA is the convergence properties dictated by the
propagator. Similar to $P^{\rm 1-loop}(k,z)$ of SPT, the term $P_{\rm
  MC}^{\rm 1-loop}(k,z)$ is roughly proportional to $P^{\rm
  L}\Delta_0^2$, but it additionally contains propagators, leading to
a large suppression at high $k$. Therefore, to the one-loop level, RPT
and CLA are expected to be most accurate at low $k$ (below the
propagator damping scale), but the predictions at low $k$ are expected
to be much accurate for RPT/CLA compared to SPT.


\section{Simulations}
\label{sec:simulations}
In this section we describe our fiducial simulations. We also run some
simulations with different settings (initial conditions, integrators
and box sizes), which are discussed in the appendix.

We begin with an explanation of our initial conditions for the
$N$-body simulations. We compute the linear power spectrum at an
initial redshift $z_{\rm ini}=31$ using {\tt CAMB}
\citep{Lewis2000}. For all the simulations in the present paper, we
adopt the standard $\Lambda$CDM model with the cosmological parameters
from the WMAP3 results \citep{Spergel2007}, $\Omega_m=0.234$,
$\Omega_\Lambda=0.766$, $\Omega_b=0.0414$, $h=0.734$, $\sigma_8=0.76$
and $n_s=0.961$, which are the current matter, cosmological constant,
baryon densities in units of the critical density, the Hubble constant
in units of $100$km$\;$s$^{-1}$Mpc$^{-1}$, density fluctuation
amplitude smoothed with a top-hat filter of radius $8h^{-1}$Mpc, and
the scalar spectral index, respectively (see table
\ref{tab:parameter}).  We then generate a linear overdensity field in
Fourier space assuming Gaussianity; the amplitude follows the Rayleigh
distribution, and the phase is uniformly distributed. We employ
$512^3$ dark matter particles within a cube of $1000h^{-1}$Mpc in each
side. We displace these particles from regular grid pre-initial
positions using the second-order Lagrangian perturbation theory (2LPT;
e.g., \cite{Crocce2006}).

We next describe the time integration scheme. We adopt a publicly
available parallel cosmological $N$-body solver {\tt Gadget2}
\citep{Springel2005}. The number of meshes used in the particle-mesh
computation is $1024^3$. We adopt a softening length of
$0.1h^{-1}$Mpc for the tree forces. We select three output redshifts;
$z=3$, $1$ and $0$, where we measure the power spectrum (see section
\ref{sec:power} for more details).

In the appendix, we discuss the convergence of our simulations against
different initial conditions, box sizes, and codes. We found that the
current setup provides a convergent result within $1\%$ of the
amplitude of the power spectrum at our scales of interest
($k\lesssim0.4h$Mpc$^{-1}$). We use this setup as our fiducial model,
and run $4$ different realizations for this model.

\begin{table}[!t]
\caption{
Adopted cosmological and simulation parameters of {\tt Gadget-2} 
for our fiducial runs.
}
\begin{center}
\begin{tabular}{|c||c|c|c|c|c|c|}
    \hline
    cosmological & $\Omega_m$ & $\Omega_\Lambda$ & $\Omega_b/\Omega_m$ & $h$ & $\sigma_8$ & $n_s$ \\
    \hline
    value & $0.234$ & $0.766$ & $0.175$ & $0.734$ & $0.76$ & $0.961$ \\
    \hline
    \hline
    simulation & box size & \# of particles
    & $z_{\rm ini}$ & \# of PM grids & softening length  
    & $N^{\rm run}$ \\
    \hline
    value & $1000h^{-1}$Mpc & $512^3$ & $31$ & $1024^3$ & $0.1h^{-1}$Mpc & $4$ \\
    \hline
\end{tabular}
\end{center}
\label{tab:parameter}
\end{table}

\section{Power Spectrum Analysis}
\label{sec:power}
\begin{figure}[!t]
\centering
\includegraphics[width=15cm]{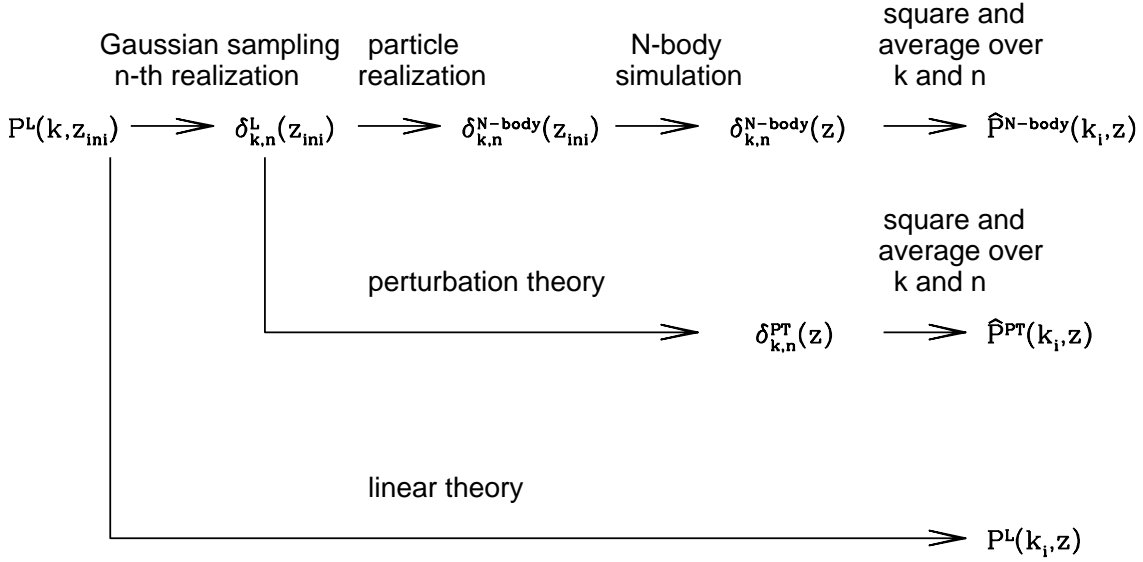}
\caption{A flow chart to illustrate our methodology to correct for the
  effect of finite box size.}
\label{fig:flowchart}
\end{figure}

\subsection{Matter density field and power spectrum in $N$-body simulations}
\label{subsec:mps}
Here we briefly describe the notation for various density fields and
power spectra which we use throughout the paper (see
figure \ref{fig:flowchart} and table \ref{tab:notation}).

\begin{table}[!t]
\begin{center}
\caption{
Notations for various density contrasts and power spectra.
}
\begin{tabular}{|c||c|c|}
    \hline
    & meaning & description \\
    \hline\hline
    $P^{\rm L}(k,z)$ & input linear power spectrum 
    & text in $\S$\ref{sec:nonlinear}\\
    \hline
    $\delta^{\rm L}_{{\bf k},n}(z)$ & density contrast 
    Gaussian-sampled from $P^{\rm L}(k,z)$ & text in 
    $\S$\ref{sec:simulations}\\
    & for the $n$-th realization &\\
    \hline
    $\delta^{N{\rm -body}}_{{\bf k},n}(z)$ & density contrast 
    realized by particles using 2LPT displacement
    & text in $\S$\ref{sec:simulations}\\
    & and evolved by $N$-body simulation & \\
    \hline
    $\hat{P}^{N{\rm -body}}(k_i,z)$ & power spectrum of the 
    $i$-th wavenumber bin estimated from $N$-body 
    & Eq.(\ref{eq:PNB})\\
    & simulations taking average over finite 
    modes and realizations & \\
    \hline
    $\delta^{\rm PT}_{{\bf k},n}(z)$ & $\delta^{\rm L}_{{\bf k},n}(z)$ 
    evolved by perturbation theory 
    & Eq.(\ref{eq:deltaPT})\\
    \hline
    $\hat{P}^{\rm PT}(k_i,z)$ & same as $\hat{P}^{N{\rm -body}}(k_i,z)$ but 
    calculated from 
    $\delta^{\rm PT}_{{\bf k},n}(z)$, not $\delta^{N{\rm -body}}_{{\bf k},n}(z)$ 
    & Eq.(\ref{eq:PPTs})\\
    \hline
    $\hat{P}^{N{\rm -body}}_{\rm corrected}(k_i,z)$ & $\hat{P}^{N{\rm -body}}(k,z)$ 
    corrected for the effect of finite volume & Eq.(\ref{eq:Pcorrect})\\
    \hline
\end{tabular}
\label{tab:notation}
\end{center}
\end{table}

We denote the Gaussian-sampled density field from $P^{\rm L}(k,z_{\rm
  ini})$ by $\delta^{\rm L}_{{\bf k},n}(z_{\rm ini})$, where the
subscript $n$ stands for the $n$-th realization. We then denote by
$\delta^{N-{\rm body}}_{{\bf k},n}(z_{\rm ini})$ the density field
{\it realized} by particles using 2LPT displacement. Note that
$\delta^{\rm L}_{{\bf k},n}(z_{\rm ini})$ is different from
$\delta^{N-{\rm body}}_{{\bf k},n}(z_{\rm ini})$; the former is
strictly Gaussian, whereas the latter slightly deviates from Gaussian
because of the nonlinearity and discreteness in the process of
displacement. The measured density contrast from the simulation output
at redshift $z$ is also expressed as $\delta^{N-{\rm body}}_{{\bf
    k},n}(z)$. In measuring $\delta^{N-{\rm body}}_{{\bf k},n}(z)$
from the simulation output, we first assign particles onto a $1024^3$
mesh using Cloud-in-Cells interpolation \citep{Hockney1981}. We then
use a FFT to calculate density contrasts in Fourier space, and divide
each Fourier mode by the Fourier transform of the window function
\citep{Angulo2008}. We made sure that the details of the interpolation
scheme and the number of mesh points do not affect the results
significantly in the scales of our interest ($k\lesssim0.4h$ Mpc
$^{-1}$). See also \citet{Jing2005} for a discussion of the effects of
aliasing.

We square $\left|\delta^{N-{\rm body}}_{{\rm k},n}(z)\right|$ and take an
average over realizations and modes: 
\begin{eqnarray}
    \hat{P}^{N-{\rm body}}(k_i,z) &\equiv& \left\langle \left|
    \delta_{{\bf k},n}^{N-{\rm body}}(z)\right|^2 
    \right\rangle_{i} \equiv \frac{1}{N^{\rm mode}_iN^{\rm run}}
    \sum_{k_i^{\rm min}<|{\bf k}|<k_i^{\rm max}}\sum_{n=1}^{N^{\rm run}} 
    \left|\delta_{{\bf k},n}^{N-{\rm body}}(z)\right|^2,\label{eq:PNB}\\
    k_i &\equiv& \frac{1}{N^{\rm mode}_i}\sum_{k_i^{\rm min}<|{\bf k}|<k_i^{\rm max}}
    \left|{\bf k}|\right.,
\label{eq:ki}
\end{eqnarray}
where $N^{\rm mode}_i$ and $N^{\rm run}$ are the numbers of modes in
the $i$-th wavenumber bin and the number of realizations, and
$k_i^{\rm min}$ and $k_i^{\rm max}$ are the minimum and the maximum
wavenumber, respectively. Note that we use $\langle...\rangle_{i}$ to
denote the average over modes in the $i$-th wavenumber bin and over
realizations: this average is not equivalent to the true ensemble
average, and the difference corresponds to the finiteness of
the simulated volume (or number of modes). In what
follows, we adopt equally-spaced bins with width $\Delta
k=0.005h$Mpc$^{-1}$.

Finally, the standard errors of the averaged power spectra of equation
(\ref{eq:PNB}) can be estimated by
\begin{eqnarray}
    \Bigl[{\rm error}\;\;{\rm of}\;\;\hat{P}^{N-{\rm body}}(k_i,z)\Bigr]^2 = 
    \frac{\left\langle\left[\left|\delta_{{\bf k},n}^{N{\rm -body}}(z)
\right|^2-\hat{P}^{N{\rm -body}}(k_i,z)\right]^2
      \right\rangle_i}{N^{\rm mode}_iN^{\rm run}}.
\label{eq:error1}
\end{eqnarray}
Note that this value indicates the uncertainty in the estimation of
the central value, not the variance among the modes in each bin.

\subsection{Corrections to the power spectrum}
\label{subsec:ICV}
The matter power spectrum measured from simulations deviates from the
prediction for the ideal ensemble average, which can be obtained only
in the limit of an infinite number of realizations or an infinite box
size. This deviation is actually important in interpreting the results
of $N$-body simulations as shown by \citet{Takahashi2008}; the matter
power spectrum measured from their $N$-body simulations does not agree
with the predictions of linear theory nor SPT even at very large
scales (e.g., $k\lesssim0.1h$Mpc$^{-1}$). They examined the effect of
a finite box size (hence a finite number of modes) and showed that the
finite-mode effect is actually responsible for the anomalous growth
rate. Here we briefly summarize their formulation of the correction.

We follow the standard notation used in cosmological perturbation
theory [see \citet{Bernardeau2002} for a review].  Let us expand the
density perturbations in $k$-space for the $n$-th $N$-body realization
as
\begin{eqnarray}
\delta^{N{\rm -body}}_{{\bf k},n}(z)=\delta_{{\bf k},n}^{\rm L}(z)
+\delta_{{\bf k},n}^{(2)}(z)+....
\end{eqnarray}
Here the second-order term is expressed as a sum of contributions from
mode couplings between two modes:
\begin{eqnarray}
\delta_{{\bf k},n}^{(2)}(z) &=& \sum_{\bf p}F^{(2)}({\bf p},{\bf k}-{\bf p};z)
\delta^{\rm L}_{{\bf p},n}(z)\delta^{\rm L}_{{\bf k}-{\bf p},n}(z),
\label{eq:delta2}
\end{eqnarray}
where the symmetrized second-order kernel $F^{(2)}$ is expressed as
\begin{eqnarray}
    F^{(2)}({\bf x},{\bf y};z)
    &=& \frac12\left[1+\epsilon(z)\right]+\frac12\frac{{\bf x}\cdot{\bf y}}{xy}
    \left(\frac xy+\frac yx\right)+\frac12\left[1-\epsilon(z)\right]
    \frac{({\bf x}\cdot{\bf y})^2}{x^2y^2},\\
    \epsilon(z) &\approx& \frac37\Omega_m^{-1/143}(z).
\end{eqnarray}
In practice, the time dependence of the kernel function is very weak,
and thus we simply set $\epsilon=3/7$ in what follows.  The power
spectrum up to the third order in $\delta^{\rm L}_{{\bf k},n}(z)$
averaged over modes in the $i$-th wavenumber bin and over realizations
is
\begin{eqnarray}
\hat{P}^{\rm PT}(k_i,z) &\equiv& \left\langle\left|
\delta_{{\bf k},n}^{\rm PT}\right|^2\right\rangle_{i},
\label{eq:PPTs}
\\
\left|\delta_{{\bf k},n}^{\rm PT}\right|^2 &\equiv& \left|
\delta_{{\bf k},n}^{\rm L}(z)\right|^2
+2\Re\left[\delta_{{\bf k},n}^{\rm L}(z)\delta_{{\bf
        -k},n}^{(2)}(z)\right],
\label{eq:deltaPT}
\end{eqnarray}
where $\Re[...]$ stands for the real part of a complex number. Though
the second term should vanish for an ensemble average over infinite
modes, it does not vanish exactly if the number of Fourier modes is
finite. The first term grows as $\propto D_+^2(z)$ while the second
term grows as $\propto D_+^3(z)$, thus the second term
becomes important at late time (i.e., at low redshifts).

The method to correct for the deviation from the ideal ensemble
average is to multiply $\hat{P}^{N-{\rm body}}(k_i,z)$ by the ratio of
$\hat{P}^{\rm PT}(k_i,z)$ and $P^{\rm L}(k_i,z)$:
\begin{eqnarray}
    \hat{P}^{\rm N-body}_{\rm corrected}(k_i,z) &\equiv& 
    \hat{P}^{\rm N-body}(k_i,z)
    \times P^{\rm L}(k_i,z)/\hat{P}^{\rm PT}(k_i,z).
\label{eq:Pcorrect}
\end{eqnarray}
The individual random nature of each $N$-body run in both
$\hat{P}^{\rm N-body}(k_i,z)$ and $\hat{P}^{\rm PT}(k_i,z)$
is weakened by this procedure [equation
(\ref{eq:Pcorrect})] as long as the predictions of perturbation theory
are sufficiently accurate. This method does not bias the
corrected value of the power spectrum even if the
perturbation theory breaks down, since $\hat{P}^{\rm PT}(k_i,z)$
approaches $P^{\rm L}(k_i,z)$ in the limit of infinite number of
Fourier modes.

As in equation (\ref{eq:error1}), the standard errors of equation
(\ref{eq:Pcorrect}) can be estimated as
\begin{eqnarray}
    \Bigl[{\rm error}\;\;{\rm of}\;\;\hat{P}^{N-{\rm body}}_{\rm corrected}
    (k_i,z)\Bigr]^2 = 
    \frac{\left\langle\left[\left|\delta_{{\bf k},n}^{N{\rm -body}}(z)
              \right|^2-\frac{\hat{P}^{N-{\rm body}}(k_i,z)
    }{\hat{P}^{\rm PT}(k_i,z)}\left|
        \delta_{{\bf k},n}^{\rm PT}(z)\right|^2 
    \right]^2
    \right\rangle_{i}}{N^{\rm run}N^{\rm mode}_i}.
\label{eq:error3}
\end{eqnarray}
The average of $\left|\delta_{{\bf k},n}^{N{\rm -body}}(z) \right|^2$
is $\hat{P}^{N-{\rm body}}(k_i,z)$ while that of $\left| \delta_{{\bf
        k},n}^{\rm PT}(z)\right|^2$ is $\hat{P}^{\rm PT}(k_i,z)$, and
thus we multiply the ratio in the second term in the numerator to
adjust the mean values.

\section{Results}
\label{sec:results}
\subsection{Comparison between $N$-body simulations and analytic models}
\label{subsec:ps}
As mentioned before, the accuracy of $N$-body simulations themselves
is not perfect and we do not regard them as perfect calibrators of
theoretical models. Instead we compare the power spectrum of our
simulations with those from theoretical predictions aiming at
determining the reliable range of wavenumbers in which both
simulations and theoretical models agree.

\begin{figure}[!ht]
\centering
\includegraphics[width=8cm]{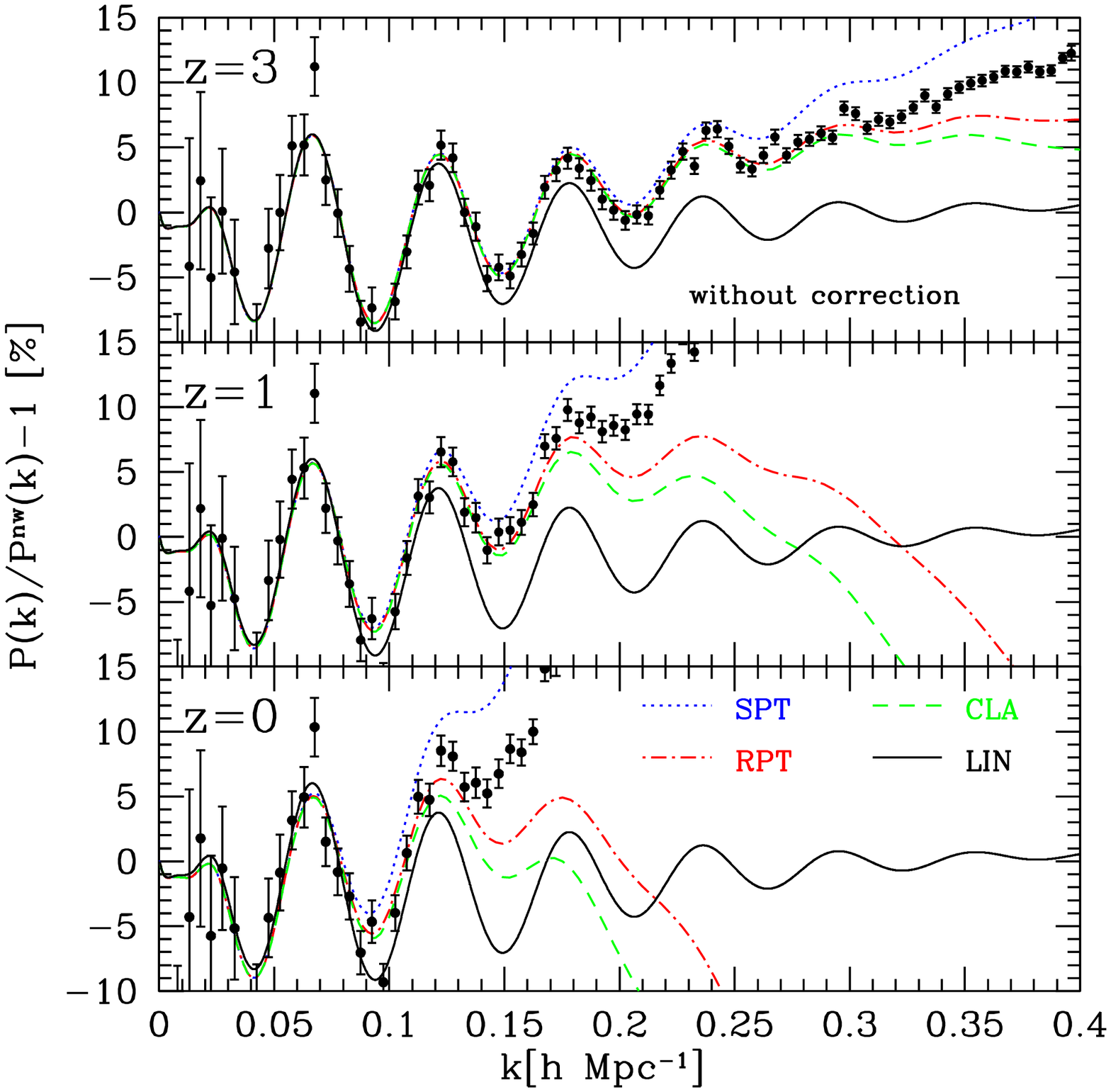}
\includegraphics[width=8cm]{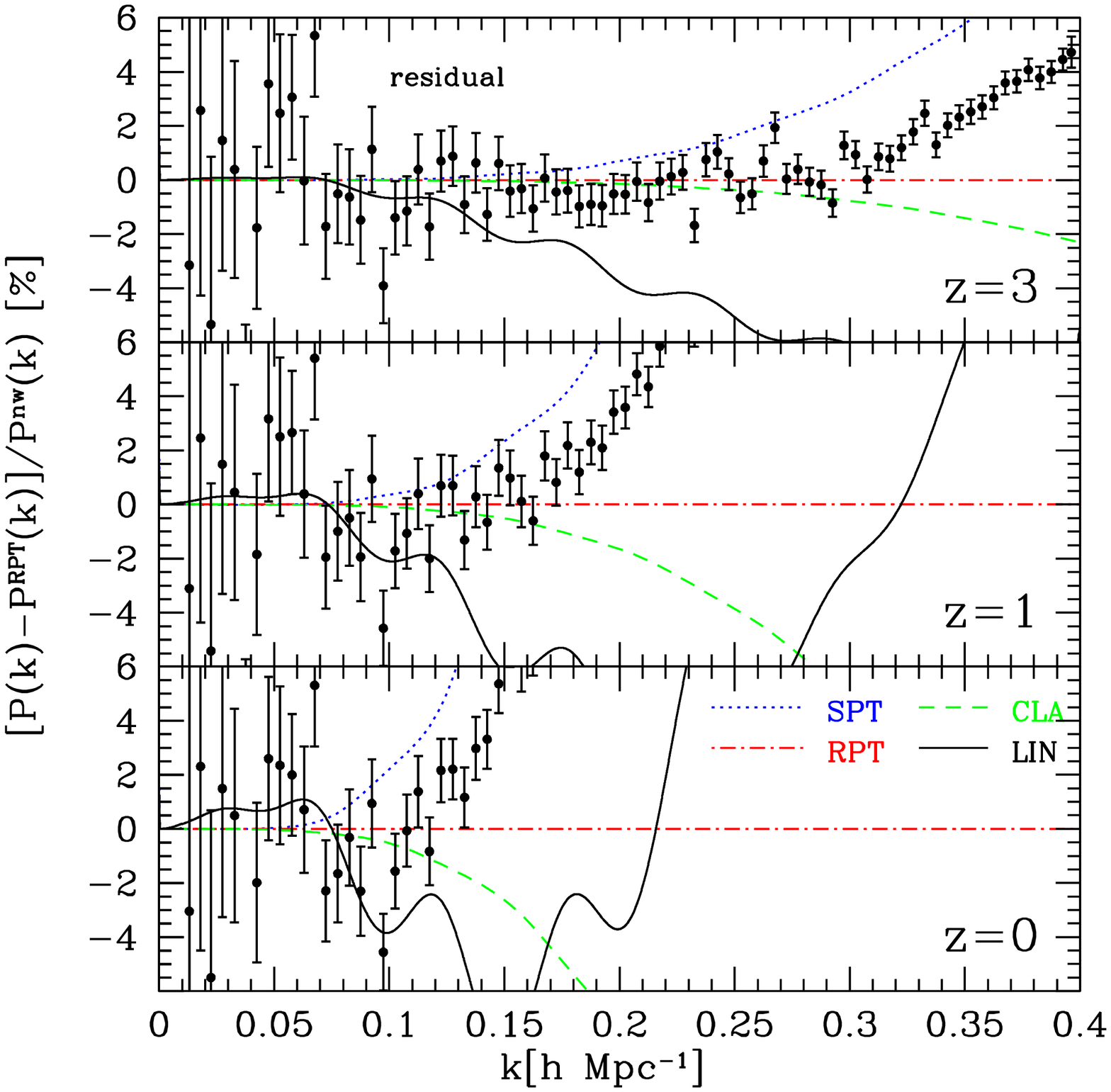}
\caption{Comparison of simulation power spectra and analytical
  predictions: {\it Left}: Power spectra of our simulations before
  correction, normalized by the no-wiggles formula; {\it top} $z=3$,
  {\it middle} $z=1$, {\it bottom} $z=0$. The errorbars show the
  standard errors [equation (\ref{eq:error1})]. Lines are theoretical
  predictions described in section \ref{sec:nonlinear}; {\it dotted}:
  standard perturbation theory (SPT), {\it dot-dashed}: renormalized
  perturbation theory (RPT), {\it dashed}: closure approximation
  (CLA), {\it solid}: linear theory (LIN). {\it Right}: Same as the
  left panel but we plot the difference from the RPT
  prediction.}
\label{fig:raw}
\end{figure}

We first compare the averaged power spectrum over the four
realizations without any corrections to the theoretical
predictions. The left panel of figure \ref{fig:raw} plots the
fractional differences from the linear power spectrum, $P^{\rm
  nw}(k,z)$, computed from no-wiggle formula of
\citet{Eisenstein1998}. The errorbars indicate the standard errors of
the estimated mean value [equation (\ref{eq:error1})]. To clarify the
differences between the $N$-body results and theoretical predictions,
we also plot the residuals from RPT in the right panel of figure
\ref{fig:raw}.

\begin{figure}[!ht]
\centering
\includegraphics[width=8cm]{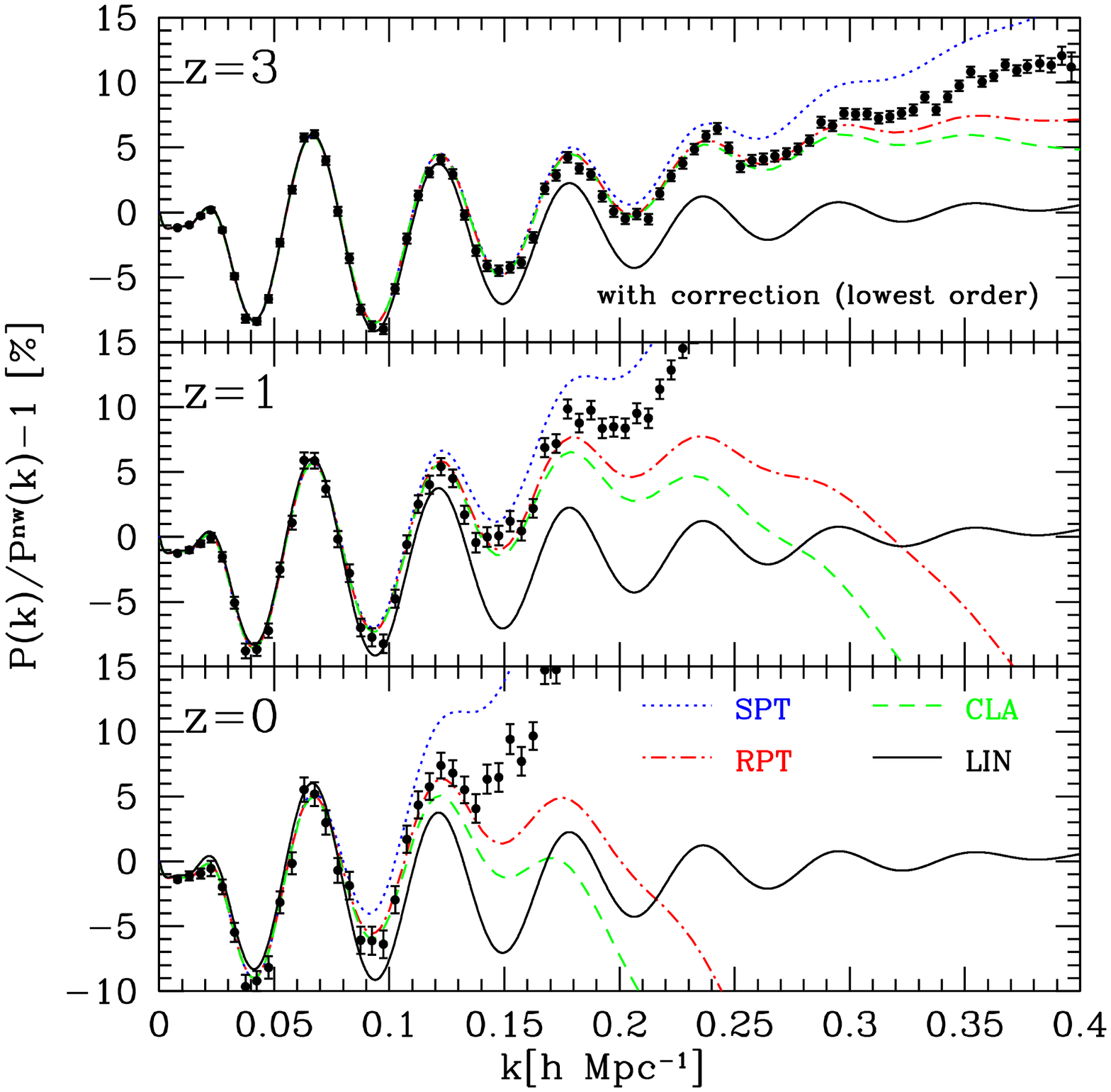}
\includegraphics[width=8cm]{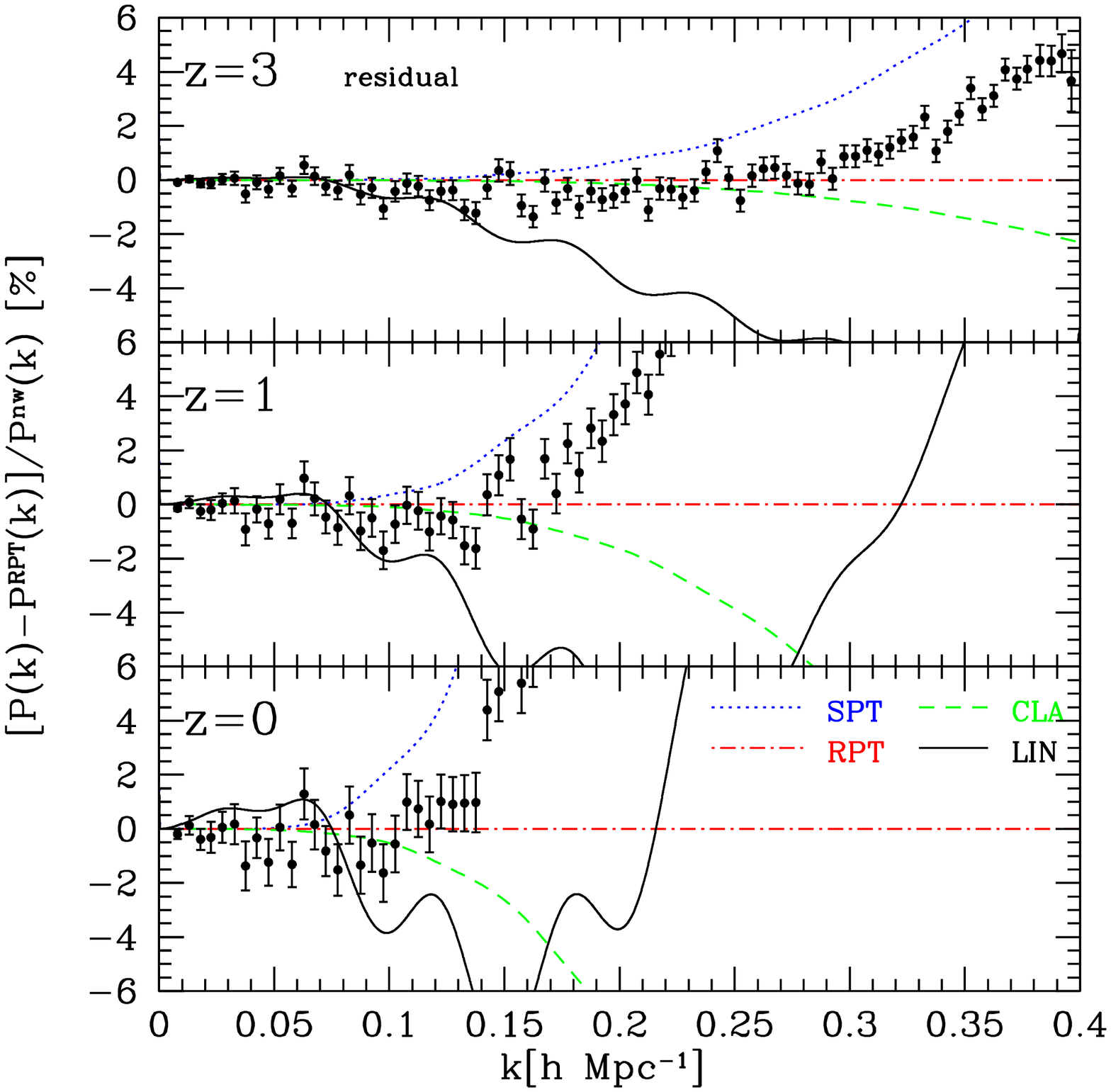}
\caption{Same as figure \ref{fig:raw}, but we correct for the
  effect of finite volume. We truncate the expansion of equation 
  (\ref{eq:deltaPT}) at the first term. The
  errorbars show equation (\ref{eq:error3}).}
\label{fig:ICV}
\end{figure}

Since large errorbars at $k\lesssim0.1h$Mpc$^{-1}$ are expected to
come mostly from the finite volume effect, we next correct for
this. In figure \ref{fig:ICV}, we plot the power spectra, based on the
procedure in $\S$ \ref{subsec:ICV}, but we truncate the expansion of
equation (\ref{eq:deltaPT}) at the first term ({\it left}: the
fractional difference from the no-wiggles formula, {\it right}:
residuals from RPT prediction).
The errorbars become significantly smaller compared with those in
figure \ref{fig:raw}, because the finiteness effect is
reduced by our methodology. Nevertheless, the results of $N$-body
simulation still exhibit a few percent error, even after the
correction.

\begin{figure}[!ht]
\centering
\includegraphics[width=8cm]{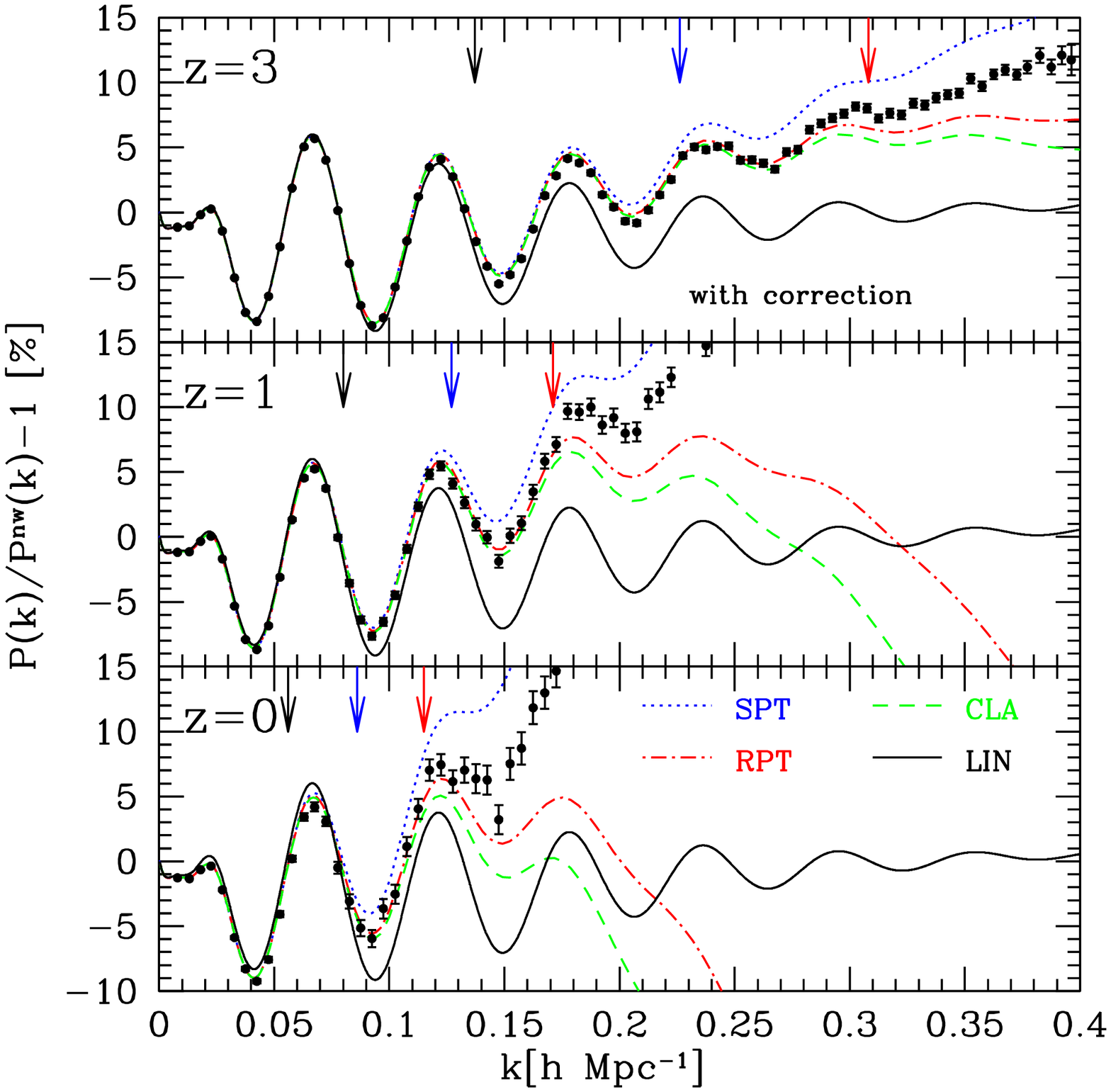}
\includegraphics[width=8cm]{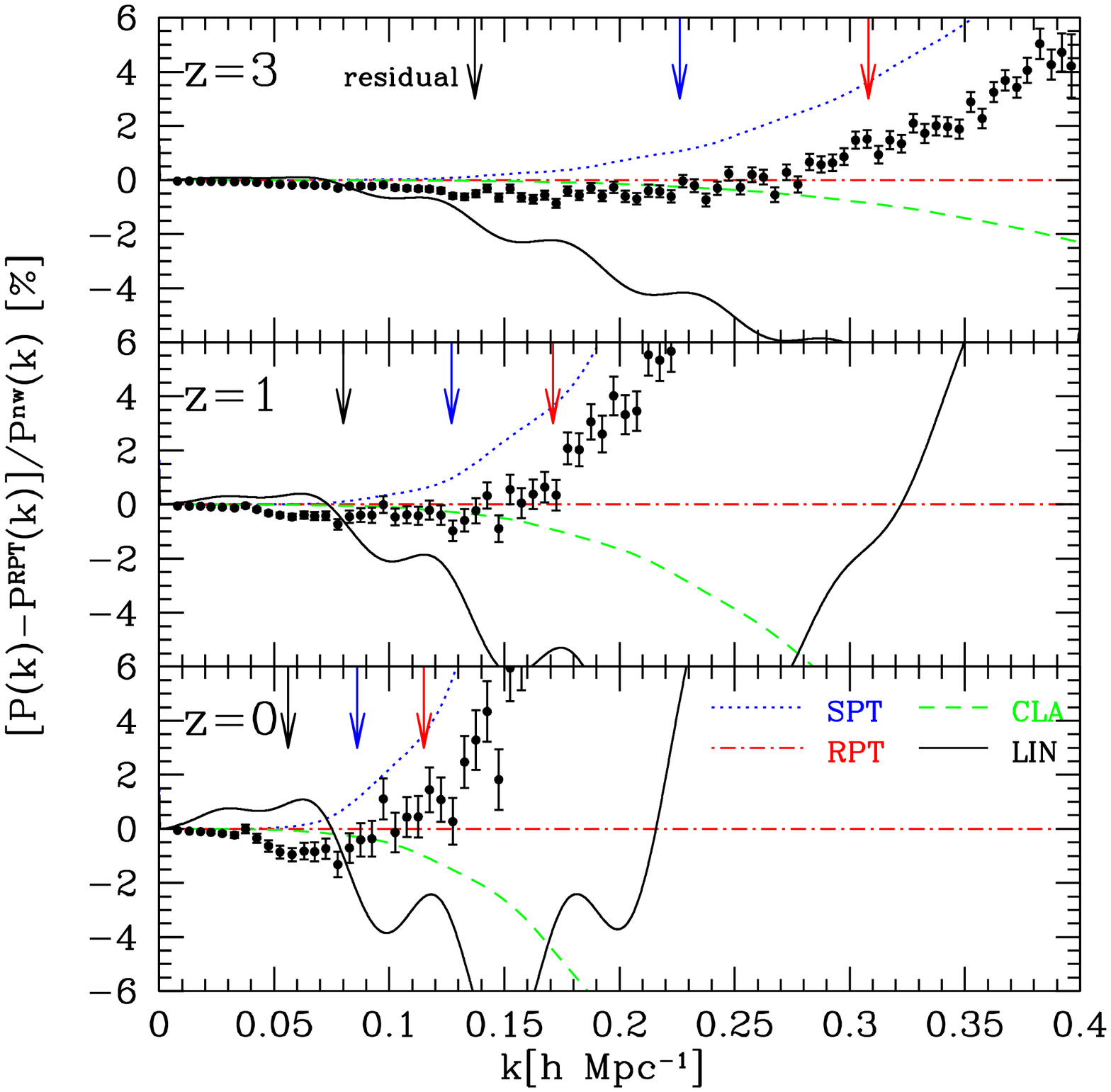}
\caption{Same as figure \ref{fig:raw}, but we correct for the effect
  of finite volume including the second term of equation
  (\ref{eq:deltaPT}). We also show the $1\%$ limit wavenumbers,
  $k_{1\%}^{\rm lim}$, for LIN, SPT and RPT/CLA by vertical arrows
  (from left to right).}
\label{fig:ICV+FMC}
\end{figure}

Next we include the second term of equation (\ref{eq:deltaPT}) for the
correction, which is coming from the mode couplings among finite
modes. Figure \ref{fig:ICV+FMC} shows the results for
  both the fractional difference from the no-wiggles formula (left),
  and the residuals from RPT predictions (right). Now the size of
errorbars are further reduced to the sub-percent level.

All the theoretical predictions plotted in figure \ref{fig:ICV+FMC}
and $N$-body simulations agree with each other well within the
errorbars at large scales up to some wavenumbers (we will determine
the range of convergence in the next subsection). Among the four
theoretical predictions, linear theory deviates from the rest at the
smallest wavenumber. The range of agreement is wider for SPT than
linear theory, as SPT includes the leading-order contribution of
nonlinear growth. RPT and CLA seem to show better agreement with
$N$-body simulations compared with SPT although all of the three
nonlinear models include their own leading-order nonlinear
corrections. This difference in the agreement ranges corresponds to
their different convergence properties; RPT and CLA show better
convergence at scales where nonlinearity is very weak.

\subsection{Convergence regime in wavenumber}
\label{subsec:valid}
We are now able to quantitatively estimate the convergence regime of
wavenumber where theories and $N$-body simulations agree. We define
two characteristic wavenumbers, $k_{1\%}^{\rm lim}$ and $k_{3\%}^{\rm
  lim}$, such that the results of $N$-body simulations and theoretical
prediction agree within $1\%$ at $k<k_{1\%}^{\rm lim} $ and within
$3\%$ at $k<k_{3\%}^{\rm lim}$, respectively. We confirmed that if we
add $1\%$ ($3\%$) Gaussian errors on the power spectra binned by
$\Delta k = 0.005h$Mpc$^{-1}$ that leads to a $\sim2\%$ ($\sim6\%$)
error on the dark energy equation of state parameter $w_{\rm DE}$ at
$z=1$ using the phase information of BAOs, which is roughly the goal
of upcoming (ongoing) surveys.

Before determining the wavenumbers, we briefly summarize three
convergence criteria previously introduced in the literature. The
first one is
\begin{eqnarray}
\Delta^2(k,z)\equiv\frac{k^3P^{\rm SPT}(k,z)}{2\pi^2} < 0.4,
\label{eq:JK}
\end{eqnarray}
introduced by \citet{Jeong2006} as a $1\%$ convergence regime of
SPT. The second one is introduced by \citet{Sefusatti2007}:
\begin{eqnarray}
    \sigma^2(R_{\rm min},z)&\equiv&\int d^3k \tilde{W}^2(qR_{\rm min})
    P^{\rm L}(q,z) = 0.25,\\
    k_{\rm max}&=&\frac{\pi}{2R_{\rm min}},
\label{eq:SK}
\end{eqnarray}
where $\tilde{W}(qR_{\rm min})$ denotes the
Fourier transform of a top-hat window function with radius $R_{\rm
  min}$. The third one is proposed by \citet{Matsubara2008}:
\begin{eqnarray}
k^2\sigma_v^2 \equiv \frac{k^2}{6\pi^2}\int_0^\infty P^{\rm L}(q,z)dq < 0.25,
\label{eq:Mat}
\end{eqnarray}
where $\sigma_v^2$ is the one-dimensional linear velocity
dispersion. This gives a scale where nonlinearity becomes important.
Note that \citet{Crocce2006b} presented similar criteria.

\begin{figure}[!ht]
\centering
\includegraphics[width=10.5cm]{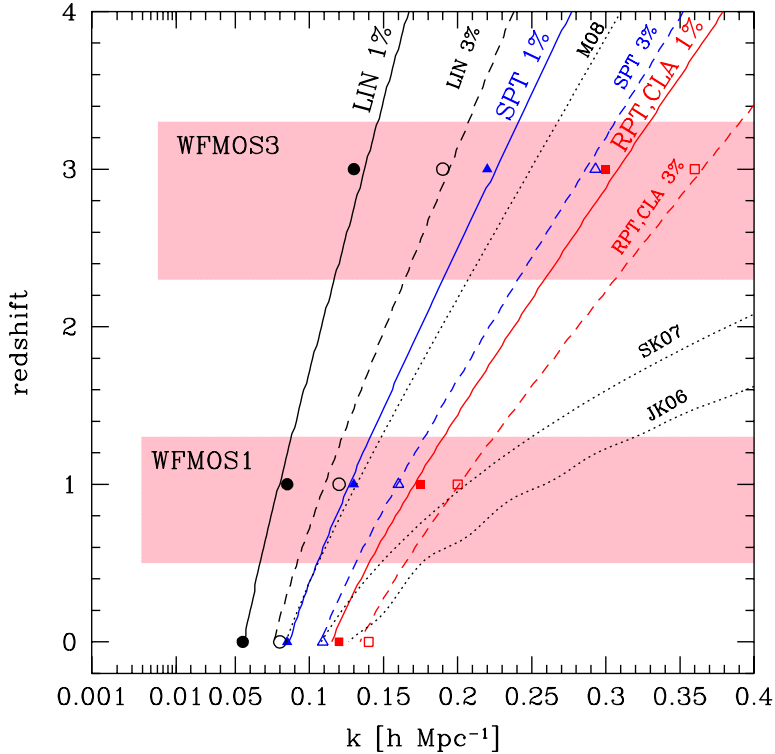}
\caption{Upper limit of reliable wavenumbers, $k_{1\%}^{\rm lim}$ and
  $k_{3\%}^{\rm lim}$, described in the text. Symbols show the values
  read from figure \ref{fig:ICV+FMC}. {\it circles}: linear theory,
  {\it triangles}: SPT, {\it squares}: RPT/CLA. Filled symbols
  correspond to $k^{\rm lim}_{1\%}$, while open ones represent $k^{\rm
    lim}_{3\%}$. The three solid lines plot equation
  (\ref{eq:validk}): $k_{1\%}^{\rm lim}$ for linear theory ($C=0.06$),
  SPT ($C=0.18$) and RPT/CLA ($C=0.35$) from left to right, and the
  dashed lines are corresponding $k_{3\%}^{\rm lim}$ ($C=0.13$, $0.3$
  and $0.5$, respectively). We also show the nonlinear wavenumbers
  proposed by \citet{Jeong2006}, \citet{Sefusatti2007} and
  \citet{Matsubara2008} as dotted lines with JK06, SK07 and M08. The
  two shaded regions show the redshift range planed by WFMOS survey
  (with minimum wavenumbers, $2\pi/V^{1/3}$. Here $V$ is the survey
  volume.).}
\label{fig:limit}
\end{figure}

We show these criteria by dotted lines against redshift in figure
\ref{fig:limit}.  The dotted curve labeled with ``JK06'' represents
the $k_{1\%}^{\rm lim}$ determined from equation (\ref{eq:JK}), and
the $k_{\rm max}$ of equation (\ref{eq:SK}) is shown by ``SK07'', and
the nonlinear scale of equation (\ref{eq:Mat}) is shown by
``M08''. Among these three criteria, the first two have similar
redshift dependence. This is because they are based on a {\it local}
value of the power spectrum or its Fourier transform convolved with a
window function. The other criterion of \citet{Matsubara2008} refers
to the integrated value of the power spectrum over scale
and thus this shows a rather mild redshift dependence of the maximum
wavenumber.

The values of $k^{\rm lim}_{1\%}$ and $k^{\rm lim}_{3\%}$ can be
directly read off from figure \ref{fig:ICV+FMC}, which are shown
in table \ref{tab:limit} and also in figure \ref{fig:limit} ($k^{\rm
  lim}_{1\%}$ by filled symbols and $k^{\rm lim}_{3\%}$ by open
symbols. {\it squares}: RPT/CLA, {\it triangles}: SPT, {\it circles}:
LIN). We empirically find that a modified version of equation
(\ref{eq:Mat}):
\begin{eqnarray}
\frac{k^2}{6\pi^2}\int_0^k P^{\rm L}(q,z)dq<C,
\label{eq:validk}
\end{eqnarray}
well reproduces the results in a unified fashion. The constant $C$ in
the right-hand-side depends on the choice of theoretical model and the
threshold value ($1\%$ or $3\%$ in this paper).  We find that $C^{\rm
  RPT,CLA}_{\rm 1\%}=0.35$, $C^{\rm SPT}_{\rm 1\%}=0.18$ and $C^{\rm
  LIN}_{\rm 1\%}=0.06$ well reproduce the $1\%$ agreement limit of
RPT/CLA, SPT and linear theory from figure \ref{fig:ICV+FMC},
respectively. We plot equation (\ref{eq:validk}) with these constant
values as solid lines in figure \ref{fig:limit} and also as vertical
arrows in figure \ref{fig:ICV+FMC}. Similarly we find that the
corresponding $3\%$ limits are given by $C^{\rm RPT,CLA}_{\rm
  3\%}=0.5$, $C^{\rm SPT}_{\rm 3\%}=0.3$ and $C^{\rm LIN}_{\rm
  3\%}=0.13$ for the three theoretical predictions (dashed lines in
figure \ref{fig:limit}).

The convergence regimes of our criteria [equation (\ref{eq:validk})]
are narrower than those previously proposed. Note, however, that the
criteria (\ref{eq:validk}) with the above constant values reasonably
describe the valid range of analytic models even at higher redshifts
($z=7$ and $15$). One of the possible sources of the difference
between our criterion and the previous ones is that they use
simulations with smaller box sizes to achieve better resolution, which
enhances a systematic error due to the finiteness of box size. More
details are discussed in the appendix.

\begin{table}[!t]
\begin{center}
    \caption{ The values of $k^{\rm lim}_{1\%}$ and $k^{\rm
        lim}_{3\%}$ read from figure \ref{fig:ICV+FMC} and the
      corresponding constant value in equation (\ref{eq:validk}).  }
\begin{tabular}{c||c c c|c||c c c|c}
& & $k^{\rm lim}_{1\%}$ [$h$Mpc$^{-1}$] & & $C_{1\%}$ & 
& $k^{\rm lim}_{3\%}$ [$h$Mpc$^{-1}$] & & $C_{3\%}$ \\
 & $z=3$ & $z=1$ & $z=0$ & & $z=3$ & $z=1$ & $z=0$ & \\
\hline
RPT/CLA & 0.3 & 0.18 & 0.12 & 0.35 & 0.36 & 0.20 & 0.14 & 0.5 \\
\hline
SPT & 0.22 & 0.13 & 0.08 & 0.18 & 0.29 & 0.16 & 0.11 & 0.3 \\
\hline
LIN & 0.13 & 0.09 & 0.06 & 0.06 & 0.19 & 0.12 & 0.08 & 0.13
\label{tab:limit}
\end{tabular}
\end{center}
\end{table}

\section{Implications for the Phases of BAOs}
\label{sec:implication}
\subsection{Extraction of the phases of BAOs from the nonlinear power
  spectra}
\label{subsec:BAO}
So far we have focused on the amplitudes of the matter power spectra,
but the phase information of BAOs is also important. This is imprinted
mainly in the leading term which is the product of the linear power
spectrum and the propagator in RPT and CLA [see equation
(\ref{eq:Pk_RPT_CLA})]. Since the other contributions, the
mode-coupling terms, are fairly smooth functions of wavenumber, the
phase information of BAOs is almost erased by the convolution of the
smoothed kernel and the propagator \footnote{The mode-coupling term
  has very small wiggly feature which comes from BAOs. This leads to a
  sub-percent level shift of the positions of ``nodes''. See
  \citet{Crocce2008} for more details.}. Therefore almost all of the
BAO information comes from the zeroth-order term in RPT/CLA. We thus
expect that the phase of BAOs predicted by $N$-body simulations and
theories agree with each other for a wider range of wavenumbers
compared to the amplitudes since the amplitudes are sensitive to
higher-order corrections.

We test the convergence of the predicted phases of BAOs as follows.
We first construct a smooth reference power spectrum using the spline
technique developed by \citet{Percival2007} and
\citet{Nishimichi2007}. We adopt a basis spline (B-spline) fitting
function as the reference spectrum, $P^{\rm B-spline}(k)$, with the
break points at $k^{\rm break}_0=0.001h$Mpc$^{-1}$ and $k^{\rm
  break}_j=(0.05j-0.025) h$Mpc$^{-1}$ where $j$ is a positive
integer. The data points to be fitted are set at $k_i$ of equation
(\ref{eq:ki}) with a bin width of $\Delta k=0.005h$Mpc$^{-1}$ for both
the $N$-body simulations results and the theoretical predictions. We
use the $N$-body power spectrum, $\hat{P}^{N-{\rm body}}_{\rm
  corrected}(k,z)$, in which the effect of finite volume is corrected
using perturbation theory. We assign equal weights to all data
points. A smooth reference spectrum is computed for the average of the
simulations and the theoretical models.

\begin{figure}[!ht]
\centering
\includegraphics[width=10.5cm]{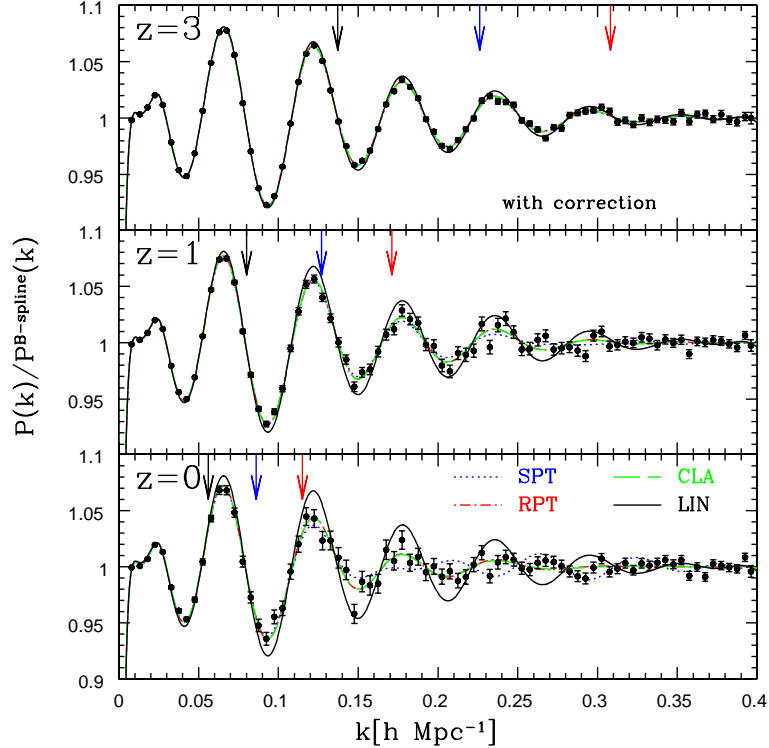}
\caption{Power spectra divided by their B-spline fits described in the
  text. Symbols, lines and vertical arrows are analogous of those in
  figure \ref{fig:ICV+FMC}.}
\label{fig:wiggle}
\end{figure}

We then divide the model and $N$-body power spectra by their
respective smooth reference power spectra, $P^{\rm B-spline}(k)$,
described above, which are shown in figure \ref{fig:wiggle}. The
symbols and lines have the same meanings as in figure
\ref{fig:ICV+FMC}. Among the three nonlinear models, RPT and CLA lie
almost on top of each other, and they are indistinguishable. These two
models show better agreement with $N$-body simulations than SPT, whose
predicted strong damping of oscillations incorrectly leads to a phase
reversal ($k\gtsim0.25h$Mpc${^{-1}}$ at $z=1$,
$k\gtsim0.16h$Mpc$^{-1}$ at $z=0$). Compared with the vertical arrows
which represent the values of $k^{\rm lim}_{1\%}$ determined by
the amplitudes of the power spectra, the
convergence regimes of BAO phases predicted by analytic models and
$N$-body simulations are quite a bit wider. We thus
conclude that the phases of BAOs are potentially more
robust and more useful in the actual analysis of BAO surveys.

\subsection{Recovery of the BAO scale from WFMOS survey}
\label{subsec:BAOfit}

With our criteria for trustable ranges of simulations and analytic
models, we are able to discuss future cosmological constraints using
BAOs. A number of simulation papers attempted to present parameter
forecasting of the dark energy equation of state parameter, $w_{\rm
  DE}\equiv p_{\rm DE}/\rho_{\rm DE}$, using BAOs as a standard ruler
(i.e., \cite{Meiksin1999}; \cite{Springeletal2005}; \cite{Angulo2005},
2008; \cite{EisensteinSW2007}; \cite{Eisenstein2007}; \cite{Seo2003},
2005; \cite{Seo2008}; \cite{Smith2007}, 2008; \cite{White2005};
\cite{Huff2007}; \cite{Jeong2006}, 2008). We follow a similar
procedure, taking into account the reliable range of wavenumbers in
the prediction of the BAO scale.

We consider the Wide-Field Multi-Object Spectrograph
(c.f., \cite{Bassett2005}, WFMOS) as a specific example in the present
analysis. We construct the template power ratio, $P/P^{\rm
  B-spline}(k)$, using RPT and linear theory as described above.  We
create mock power spectra by adding Gaussian random errors
\citep{Feldman1994}:
\begin{eqnarray}
    \Delta\left(\frac{P(k)}{P^{\rm B-spline}(k)}\right) \approx 
    \frac{\Delta P(k)}{P(k)}
    = \sqrt{\frac{2}{N^{\rm modes}}}\left[1+\frac{1}{n_gP(k)}\right].
\label{eq:error}
\end{eqnarray}
We neglect the errors arising from the construction process of
$P^{\rm B-spline}(k)$. In the above expression, $N^{\rm modes}$ is
the number of modes in the wavenumber bin $k\sim k+\Delta k$:
\begin{eqnarray}
N^{\rm modes}= \frac{k^2\Delta kV}{2\pi^2},
\end{eqnarray}
where $V$ denotes the survey volume, and $n_g$ in equation
(\ref{eq:error}) is the mean number density of galaxies. We set
$\Delta k$ to be very small ($0.001h$Mpc$^{-1}$) so as not to affect
the results and use the nonlinear matter power spectrum of
\citet{Smith2003} for the calculation of the shot noise contribution,
$[n_gP(k)]^{-1}$, just for simplicity.

We fit these mock data to the corresponding theoretical template
$P/P^{\rm B-spline}(\alpha k)$ with a single free parameter $\alpha$,
the ratio of the true distance scale to the assumed one. The
response of the {\it observed} power spectrum, $P_{\rm obs}(k,z)$, to
the parameter $\alpha$ can be written as (e.g., \cite{Jeong2008})
\begin{eqnarray}
    P_{\rm obs}(k,z) = \alpha^{-3}P_{\rm true}(\alpha k,z);\qquad
    \alpha \equiv \frac{D_{\rm V, true}}{D_{\rm V, fid}},
\end{eqnarray}
where $P_{\rm true}(k,z)$ is the true power spectrum and $D_{\rm
  V}(z)\propto [D_{\rm A}^2(z)/H(z)]^{1/3}$, where $D_A(z)$ is the
angular diameter distance \citep{Eisenstein2005}. $D_{\rm V, true}(z)$
represents the true distance, while $D_{\rm V, fid}(z)$ is calculated
assuming the fiducial cosmology. In fitting the templates, we fix the
value of the minimum wavenumber, $k_{\rm min}$, to be
$0.02h$Mpc$^{-1}$, while the value of the maximum wavenumber, $k_{\rm
  max}$, is varied. The assumed parameters for the $z\sim1$ ($z\sim3$)
WFMOS survey are taken from \citet{Glazebrook2005}: $V=4h^{-3}$Gpc$^3$
($1h^{-3}$Gpc$^3$) for the volume, $n_g=5\times 10^{-4}h^3$Mpc$^{-3}$
($5\times 10^{-4}h^3$Mpc$^{-3}$) for the galaxy number density, and
$b=2.0$ ($3.5$) for the linear bias parameter.

\begin{figure}[!ht]
\centering
\includegraphics[width=10.5cm]{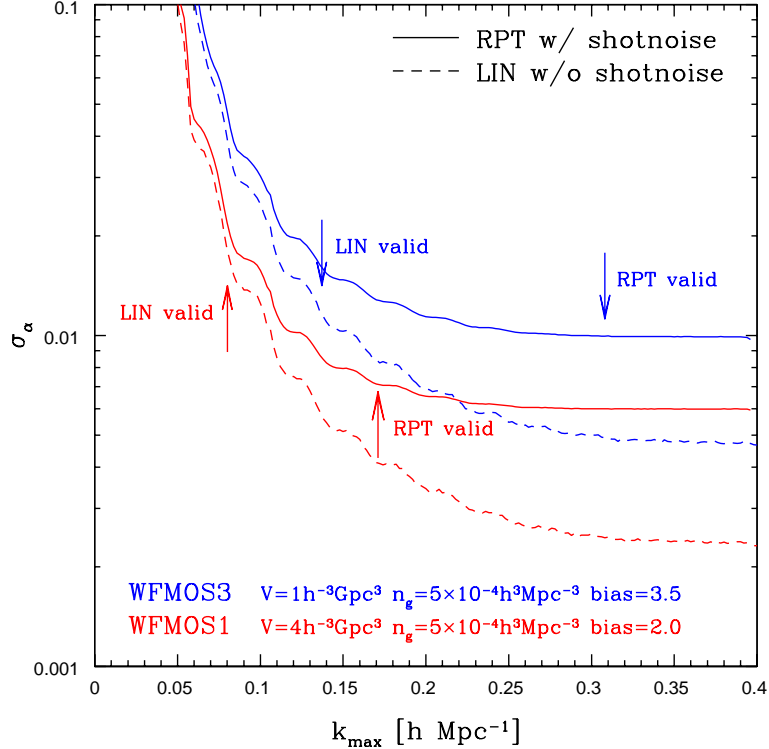}
\caption{Estimated constraints on the scale shift parameter, $\alpha$,
  for both the wide ($z\sim1$, red) and the deep ($z\sim3$, blue)
  components of the proposed WFMOS survey. The solid lines show the
  results when we use RPT as the template spectrum, while the dashed
  ones correspond to an ideal case of no shot noise and no nonlinear
  degradation of BAOs (using the linear theory
  template). We fix the minimum wavenumber to be used in the fitting,
  $k_{\rm min} = 0.02h$Mpc$^{-1}$, and the results are plotted as a
  function of the maximum wavenumber. The assumed survey parameters
  are written in the figure.}
\label{fig:sigma_alpha}
\end{figure}

We employ a Markov chain Monte Carlo fitting routine to determine the
uncertainty in the scale shift, $\sigma_\alpha$, adopting RPT for the
fiducial spectrum. We plan to do a similar fitting with more free
parameters in future work. Figure \ref{fig:sigma_alpha} show the
uncertainty in the scale shift, $\sigma_\alpha$, as a function of
$k_{\rm max}$ (solid lines). We also plot the results when we do not
include shot noise and nonlinear degradation of BAOs (linear template)
by the dashed lines. As a reference, vertical solid and dashed arrows
indicate $k^{\rm lim}_{1\%}$ for RPT and linear theory,
respectively. The difference of solid and dashed lines correspond to
the impact of nonlinearity and shot noise, i.e., the information of
phase of BAOs is lost due to these effects. The resulting error of
$\alpha$ is $0.7(1.0)\%$ at $z=1(3)$ when we adopt $k^{\rm lim}_{1\%}$
of RPT/CLA as $k_{\rm max}$. These values are slightly improved and
become $0.6(1.0)\%$ if we increase $k_{\rm max}$ to $0.4h$Mpc$^{-1}$.
We can put a strong constraint on $\alpha$ even if we do not include
the modes at $k>k_{1\%}^{\rm lim}$ of RPT in the analysis. In reality,
however, we need to take account of {\it galaxy} clustering in {\it
  redshift space} in addition to what we did in the current paper,
which will be presented in future work.

\section{Summary}
\label{sec:summary}
We have carried out a systematic comparison of power
spectra calculated from $N$-body simulations and various theoretical
models. We correct for the effect of finite volume of $N$-body
simulations using the perturbation-theory-based methodology developed
in \citet{Takahashi2008}.

We find that our simulations agree with all the theoretical models
used in the present paper at large length scales
($k\lesssim0.1h$Mpc$^{-1}$). At smaller length scales where
nonlinearity is mild, our simulations, RPT (Crocce \&
Scoccimarro 2006a, b, 2008) and CLA \citep{Taruya2008} agree with
$1\%$ and $3\%$ accuracies up to $k^{\rm lim}_{1\%}$ and $k^{\rm
  lim}_{3\%}$ as presented in figure \ref{fig:limit}.  These
convergence regimes depend on the redshift and can be explained by a
simple empirical formula of equation (\ref{eq:validk}).  We also
showed that the phase information of BAOs extracted from power spectra
using B-spline fitting is more robust than the amplitudes: predicted
phases of nonlinear theories and $N$-body simulations
agree well in a wider range of wavenumbers than the
amplitudes of the power spectrum.

The currently achieved accuracies of both theoretical predictions and
simulations are sufficient to interpret the data from future surveys
like WFMOS: one can put a tight constraint on the BAO scale
($\sigma_\alpha\lesssim1\%$) using modes within our convergence
regimes. There is the potential to extract more than simply the BAO
scale from the power spectrum, such as the sum of the neutrino masses,
if one is able to exploit smaller scale information (e.g.,
\cite{Saito2008}). To this end, it is important to consider
higher-order corrections (i.e., 2-loop terms) on the theory end and
higher resolution simulations. It is also important to investigate the
accuracy of the velocity field in $N$-body simulations to model
accurately in redshift space.  These are future investigations and a
study in redshift space is now in progress.

We also checked the convergence of the matter power spectrum in
$N$-body simulations changing the initial conditions, $N$-body
solvers, and box sizes. Among these we showed that $N$-body
simulations with small box sizes ($\lesssim500h^{-1}$Mpc) suffer from
systematic enhancements of the matter power spectra at BAO scales
(e.g., \cite{Springeletal2005, Seo2005, Jeong2006}).

\bigskip

We thank S.~Habib, R.~Scoccimarro, M.~Takada and E.~Reese for useful
suggestions and comments on the present paper. This work is supported
in part by Japan Society for Promotion of Science (JSPS) Core-to-Core
Program ``International Research Network for Dark Energy''. T.N.,
A.S., K.Yahata and I.K. acknowledge the support from the JSPS Research
Fellows. S.S. is supported by Global COE Program "the Physical
Sciences Frontier", MEXT, Japan. Y.P.J. is supported by NSFC
(10533030, 10821302, 10878001), by the Knowledge Innovation Program of
CAS (No. KJCX2-YW-T05), and by 973 Program (No.2007CB815402). This
work is supported by Grants-in-Aid for Scientific Research from the
JSPS (Nos.~18740132,~18654047,~19840008) and by Grant-in-Aid for
Scientific Research on Priority Areas No. 467 "Probing the Dark Energy
through an Extremely Wide \& Deep Survey with Subaru Telescope".

\appendix

\bigskip


\appendix
\section{Convergence check of different $N$-body codes and dependence on simulation parameters}
\label{sec:convergence}
\subsection{Initial Condition Generator}
\label{subsec:IC}

We test two methods for generating cosmological initial conditions.
One is the Zel'dovich approximation (\cite{Zeldovich1970}; hereafter
ZA) commonly used for cosmological simulations, and the other is
second-order Lagrangian perturbation theory (2LPT; e.g.,
\cite{Crocce2006}). Starting from the same linear density field,
$\delta^{\rm L}_{{\bf k},n}$, we generate two initial conditions at
$z_{\rm ini}=31$ from $N$-body runs using ZA and 2LPT. Then we compare
the power spectra evolved with {\tt Gadget2}.

\begin{figure}
\centering
\includegraphics[width=10.5cm]{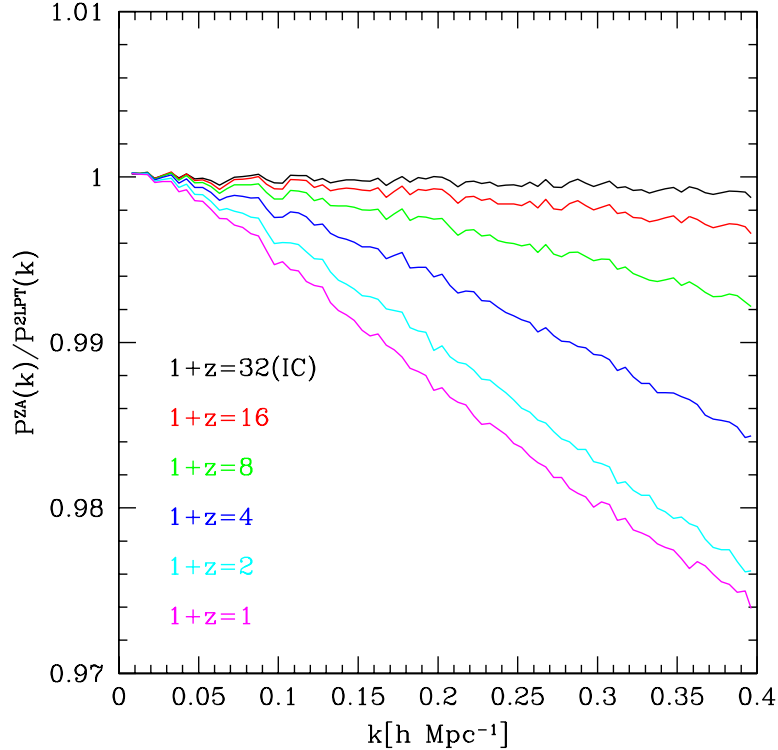}
\caption{We plot the evolution of the power spectrum from two
  simulations using a ZA initial condition and a 2LPT. To see the
  difference clearly, we show the ratio $P^{\rm ZA}/P^{\rm 2LPT}$ at a
  number of output times.}
\label{fig:ZA2LPT}
\end{figure}

Figure \ref{fig:ZA2LPT} shows the ratio of the two matter power
spectra from simulations at various output redshifts. The overall
trend that $P^{\rm ZA}(k)$ has smaller amplitudes at all scales than
$P^{\rm 2LPT}(k)$ is consistent with the result of \cite{Crocce2006}.
This is because the higher-order terms in ZA do not necessarily
correspond to the growing solutions. The plot indicates that
differences in the initial conditions affect the power spectrum at the
few percent level. We confirmed that this difference between 2LPT and
ZA reduces to subpercent when we increase the starting redshift to
$z_{\rm ini}=127$. However, we need a very large number of mesh points
in the calculation of the PM force to avoid large errors introduced by
the tree force that arise when the density field is nearly
uniform. This is very time consuming. Thus we need 2LPT initial
condition for the purpose of present analysis if we start the
simulations at $z_{\rm ini}=31$. We also confirm that our 2LPT result
is not affected when we change the starting redshift to $z_{\rm
  ini}=63$ or $z_{\rm ini}=127$.

\subsection{Comparison using different  $N$-body solvers}
\label{subsec:code}
Next we compare three different N-body solvers; a Tree-Particle-Mesh
(Tree-PM) solver {\tt Gadget2} \citep{Springel2005}, another Tree-PM
code {\tt TPM} of \citet{Bode2003}, and a
Particle-Particle-Particle-Mesh (P$^3$M) code developed in Jing \&
Suto (1998, 2002; hereafter {\tt JS02}). We use identical initial
conditions as that of one of our fiducial runs, $512^3$ particles in a
box of $1000h^{-1}$Mpc. We stop the simulations at $z=3$. In addition
to the Tree-PM and P$^3$M calculations, we also run simulations using
only the PM part of the three codes.  We measure the power spectra for
the simulation output at $z=3$ and correct for the effect of finite
volume (see section \ref{sec:power} above) so that the comparison is
clearer.

\begin{figure}
\centering
\includegraphics[width=10.5cm]{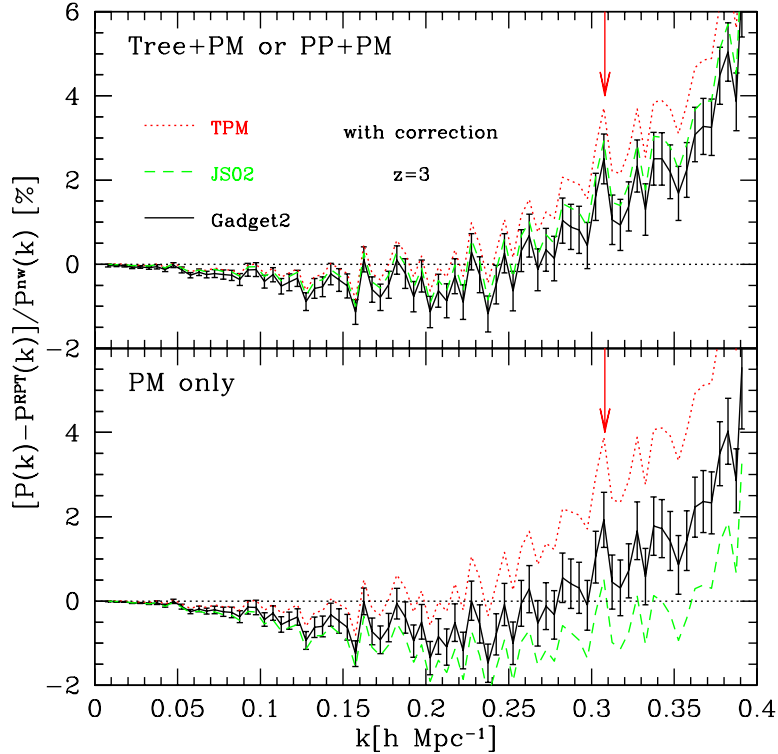}
\caption{A comparison of different $N$-body simulation codes. We plot
  $[\hat{P}^{N-{\rm body}}_{\rm corrected}(k)-P^{\rm RPT}(k)]/P^{\rm
    nw}(k)$ at $z=3$. Different line types correspond to {\tt TPM}
  (dotted), {\tt JS02} (dashed) and {\tt Gadget2} (solid),
  respectively. The vertical arrows show the limit wavenumber $k^{\rm
    lim}_{1\%}$ for RPT/CLA. The errorbars are given by equation
  (\ref{eq:error3}) for {\tt Gadget2}. {\it Upper}: Results of Tree-PM
  and PP-PM calculations. {\it Lower}: Results of PM only
  calculations.  }
\label{fig:comp}
\end{figure}

Figure \ref{fig:comp} shows the relative difference from the RPT
prediction, $[\hat{P}^{N-{\rm body}}_{\rm corrected}(k)-P^{\rm
  RPT}(k)]/P^{\rm nw}(k)$. The upper panel shows the results when we
include the short-range forces (i.e., tree or particle-particle
force). The result of {\tt TPM}, {\tt JS02} and {\tt Gadget2} are
shown by dotted, dashed and solid lines with errorbars,
respectively. We omit the errorbars for {\tt TPM} and {\tt JS02} as
these are almost the same as for {\tt Gadget2}, which is given by
equation (\ref{eq:error3}). The overall behaviours for the three codes
are very consistent, and the difference is important ($\gtsim1\%$)
only at smaller scale than $k^{\rm lim}_{1\%}$ of RPT/CLA shown by the
vertical arrow.  In the bottom panel we plot the results of the three
codes when we use only PM forces. The degree of dispersion among the
three is larger than that in the upper panel, and is $\sim3\%$ at
$k^{\rm lim}_{1\%}$ of RPT/CLA. The short-range force is helpful to
increase the accuracy at this level at this scale. We conclude that
our main result is not affected so much by the choice of particular
$N$-body solver when we turn on the short-range force.

\subsection{Effects of finite box size and resolution}
\label{subsec:boxsize}

The finite box size introduces an additional variance in the power
spectrum as shown in \citet{Takahashi2008} and this work. It is
possible that this finiteness affects not only the variance but also
the mean value. The resolution of simulations may also affect the
results. We thus investigate the box size/resolution dependence of the
power spectra from $N$-body simulations.

\begin{figure}[!ht]
\centering
\includegraphics[width=7.5cm]{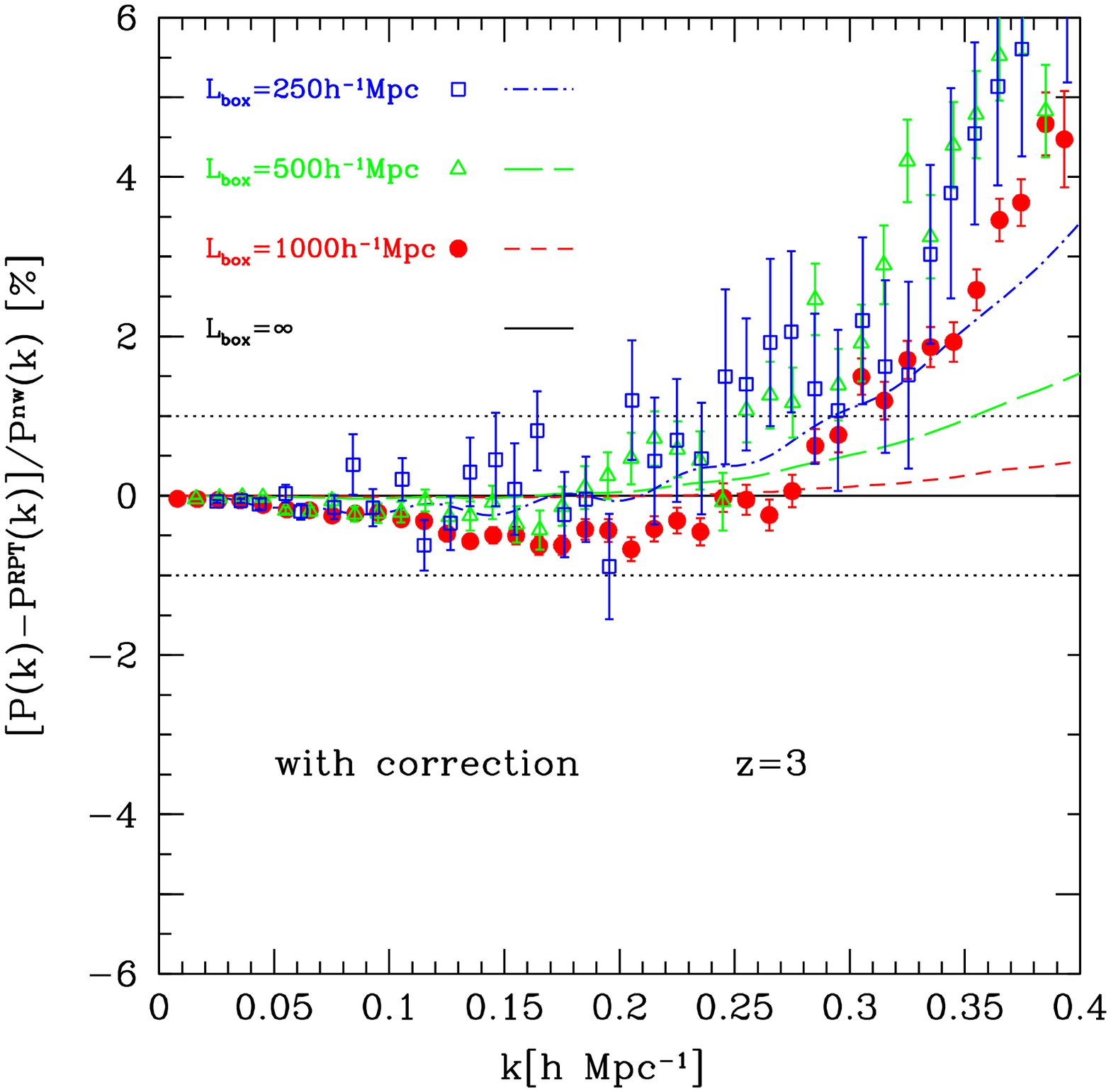}
\includegraphics[width=7.5cm]{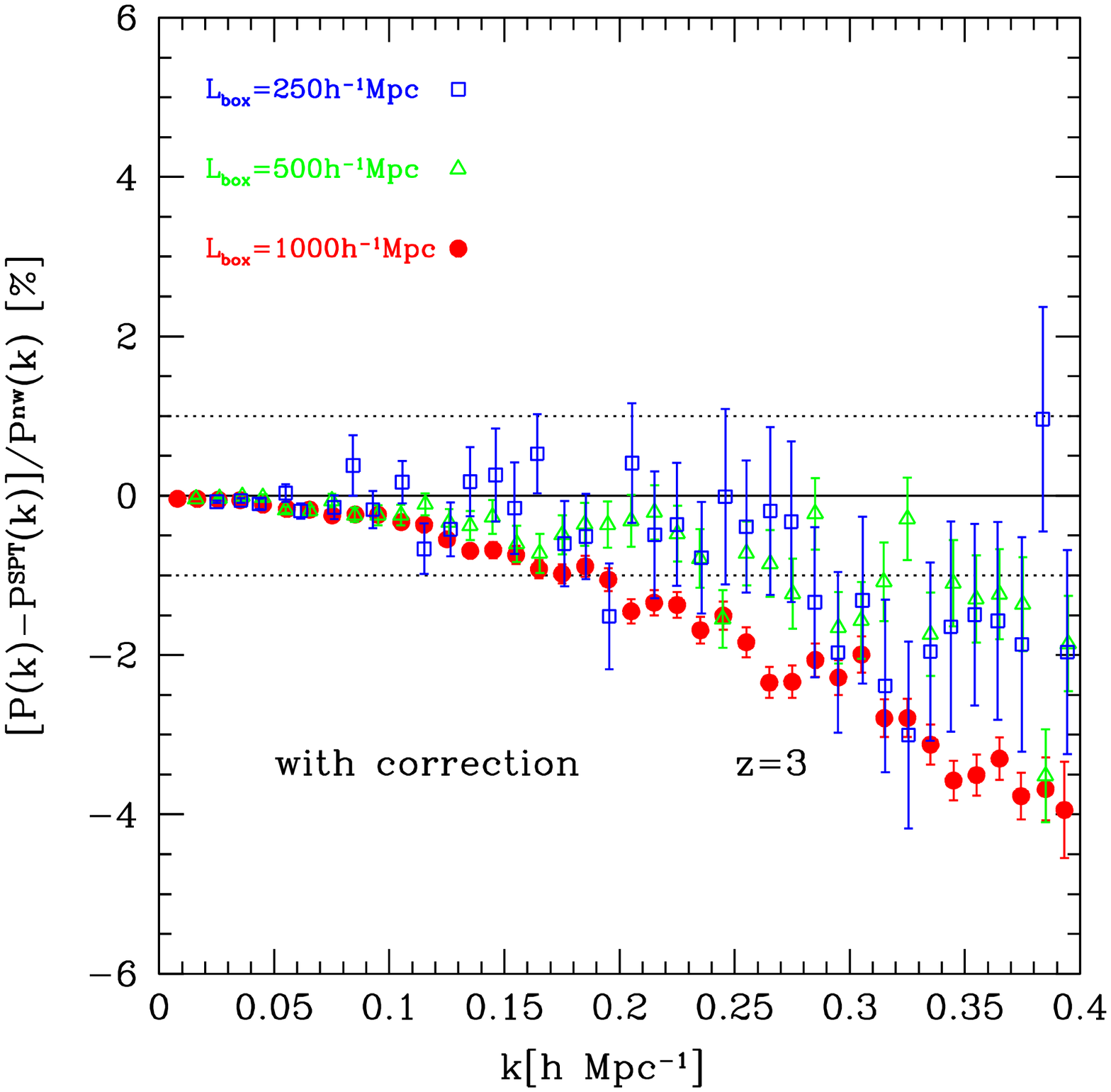}
\caption{We compare results of $N$-body simulations for different box
  sizes at $z=3$. {\it circle}: $1000h^{-1}$Mpc, {\it triangle}:
  $500h^{-1}$Mpc, {\it square}: $250h^{-1}$Mpc. {\it Left}: comparison
  with the RPT prediction (dotted line). We also plot predictions of
  RPT taking account of the finiteness of the volume by introducing a
  cut off scale: {\it solid}: {\it short-dashed}: $1000h^{-1}$Mpc
  (dashed), $500h^{-1}$Mpc (long-dashed) and $250h^{-1}$Mpc
  (dot-dashed). {\it Right}: comparison with the SPT prediction, in
  which we do not take account of the finiteness effect.}
\label{fig:boxsize}
\end{figure}

In order to increase the spatial resolution, we run simulations
varying the box size [$1000$ ($4$ realizations; fiducial), $500$ ($7$
realizations) and $250h^{-1}$Mpc ($9$ realizations)] but keeping the
number of particles to be the same as our fiducial runs ($512^3$) to
check the convergence. For this test, we use outputs at $z=3$. The
results are shown in figure \ref{fig:boxsize}.  Symbols correspond to
$L_{\rm box}=1000h^{-1}$Mpc (circle), $500h^{-1}$Mpc (triangle) and
$250h^{-1}$Mpc (square). We adopt the bin width $\Delta
k=0.01h$Mpc$^{-1}$ in this section for clarity. We also calculate the
RPT predictions with the effect of finite box size by introducing a
cut-off scale ($k_{\rm box}\equiv2\pi/L_{\rm box}$) in the
integration, which are shown in the left panel. The solid line is the
normal RPT result (without cut-off), and the other lines take account
of the effect of finite box size (short-dashed: $1000h^{-1}$Mpc,
long-dashed: $500h^{-1}$Mpc and dot-dashed: $250h^{-1}$Mpc). Note that
we show only the results without cutoff in the text. In the right
panel, we plot a comparison with the SPT prediction (no cut-off) to
show the dependence on the box size of the valid ranges of SPT.

In the left panel, one may notice the difference of theoretical lines
at large wavenumbers. This reflects the finiteness of the box:
short-dashed line ($L_{\rm box} = 1000h^{-1}$ Mpc case) deviates from
solid line (infinite volume case) only at $ k > 0.3h$ Mpc $^{-1}$ and
is very small ($\sim0.5\%$ at $k=0.4h$ Mpc $^{-1}$), whereas
long-dashed and dash-dotted ones ($L_{\rm box} = 500$ and $250h^{-1}$
Mpc cases) diverge at smaller wavenumbers and to a greater degree
($\sim1.5\%$ and $\sim3\%$ at $k=0.4h$ Mpc $^{-1}$,
respectively). Although the results of $N$-body simulations with
smaller box sizes have larger variances due to smaller total volumes
than the fiducial runs, box size dependence appears at small
scales. The runs with smaller box size tend to have larger amplitudes,
which is consistent with RPT predictions with cutoff scales. We thus
conclude that the results of simulations also suffer from a systematic
effect due to finite box size, and adopt $L_{\rm box} =
1000h^{-1}$Mpc for our fiducial model, where the effect is
$\lesssim0.5\%$.

Next, the convergence regime of SPT seems wider if one believes the
results of the smaller box size runs in the right panel. We consider
these wider valid ranges are just coincidences due to the systematic
effects in the simulations with smaller box sizes and should not be
trusted. One should keep this in mind when one uses the results of
simulations with small box size. Even the Millennium simulation
\citep{Springeletal2005} has only $500h^{-1}$Mpc on each side,
although it has many more particles ($2160^3$) than ours and thus the
spacial resolution is better. This box size introduces $\gtsim1\%$
systematics at $k\gtsim0.35h$Mpc$^{-1}$ according to our figure
\ref{fig:boxsize}. Part of the disagreement in the valid range of SPT
between our results and that of \citet{Jeong2006} may be due to this
effect, as they use $L_{\rm box}=512$, $256$, $128$ and $64h^{-1}$Mpc.

\end{document}